\documentclass[sn-standardnature]{sn-jnl} 
\usepackage[resetlabels]{multibib}
\newcites{supp}{Supplementary References}
\usepackage{bigints}
\jyear{2021}%
\theoremstyle{thmstyleone}%
\theoremstyle{thmstyletwo}%
\theoremstyle{thmstylethree}%
\usepackage{textgreek}
\usepackage[utf8]{inputenc}
\usepackage{charter}
\usepackage{amsmath}
\usepackage{amssymb}
\usepackage{amsthm}
\usepackage{mathtools}
\usepackage{braket}
\usepackage{soul}
\usepackage{anyfontsize}
\usepackage[mathlines]{lineno}

\begin{document}

\title[Article]{Reversible tuning of nanowire quantum dot to atomic transitions}

\author*[1,2]{\fnm{Rubayet} \sur{Al Maruf}}\email{ralmaruf@uwaterloo.ca}
\author[1,2]{\fnm{Sreesh} \sur{Venuturumilli}}\email{ssvenuturumilli@uwaterloo.ca}
\author[1,2]{\fnm{Divya} \sur{Bharadwaj}}\email{divyabharadwaj4@gmail.com}
\author[1,3]{\fnm{Paul} \sur{Anderson}}\email{p8anderson@uwaterloo.ca}
\author[1,3]{\fnm{Jiawei} \sur{Qiu}}\email{jiawei.qiu@uwaterloo.ca}
\author[1,2]{\fnm{Yujia} \sur{Yuan}}\email{y246yuan@uwaterloo.ca}
\author[1,2]{\fnm{Mohd} \sur{Zeeshan}}\email{m5zeeshan@uwaterloo.ca}
\author[1,2]{\fnm{Behrooz} \sur{Semnani}}\email{bsemnani@uwaterloo.ca}
\author[4]{\fnm{Philip} \sur{J. Poole}}\email{philip.poole@nrc-cnrc.gc.ca}
\author[4]{\fnm{Dan} \sur{Dalacu}}\email{dan.dalacu@nrc-cnrc.gc.ca}
\author[1,3]{\fnm{Kevin} \sur{Resch}}\email{kresch@uwaterloo.ca}
\author[1,2]{\fnm{Michael} \sur{E. Reimer}}\email{michael.reimer@uwaterloo.ca}
\author[1,2]{\fnm{Michal} \sur{Bajcsy}}\email{mbajcsy@uwaterloo.ca}

\affil[1]{\orgdiv{Institute for Quantum Computing}, \orgname{University of Waterloo}, \orgaddress{ \city{Waterloo}, \postcode{N2L 3G1}, \state{Ontario}, \country{Canada}}}

\affil[2]{\orgdiv{Electrical and Computer Engineering}, \orgname{University of Waterloo}, \orgaddress{\city{Waterloo}, \postcode{N2L 3G1}, \state{Ontario}, \country{Canada}}}

\affil[3]{\orgdiv{Dept. of Physics and Astronomy}, \orgname{University of Waterloo}, \orgaddress{\city{Waterloo}, \postcode{N2L 3G1}, \state{Ontario}, \country{Canada}}}

\affil[4]{\orgname{National Research Council of Canada}, \orgaddress{\city{Ottawa}, \postcode{K1A 0R6}, \state{Ontario}, \country{Canada}}}

\abstract{
Quantum dots embedded in semiconductor photonic nanowires (NW-QDs) can deterministically produce single-photons and entangled photon pairs at high repetition rates. These photons can be efficiently coupled from the photonic nanowire into free space or optical fibers thanks to the sharp tip of the nanowire, which provides impedance matching. However, precise control of the NW-QD emission frequency in a way that is reversible, does not degrade the properties of the emitted photons, and can be used independently for individual NW-QDs on the same chip has so far remained a challenge.  Resolving this issue is crucial for applications when interfacing the photons with quantum systems that require MHz to sub-GHz precision, such as atomic ensembles acting as memories in a quantum network. Here, we demonstrate a reversible tuning method that can tune the emission frequency of a NW-QD by more than 300 GHz with sub-GHz precision. We achieve this through gas condensation that is then partially reversed with localized laser ablation. This process finely adjusts stress applied to the quantum dots, thereby tuning their emission frequency. We validate the precision and stability of this method by tuning the frequency of the emitted single-photons across an atomic resonance to probe its absorption and dispersion. We observed up to 80\% absorption of the single-photons from NW-QD in hot caesium vapour at the D\textsubscript{1}-line resonances and a 75-fold decrease in group velocity associated with the hyperfine transitions of the D\textsubscript{1}-line ground states. We observed no discernible effects  in the second-order autocorrelation function, lifetime, or linewidth of the NW-QD emission for up to 300 GHz of tuning and we saw minimal effects on the fine structure splitting of the NW-QD when tuning up to 100 GHz.
}


\maketitle
\section*{Introduction}\label{Introduction}
Epitaxially grown semiconductor quantum dots (QD) are among the most promising candidates for an ideal solid-state single-photon source. This is due to the purity of the single-photon states that QDs can emit, the deterministic nature of the emission following an input trigger which can be applied with high repetition rate, and indistinguishability of the emitted photons in terms of polarization, spatial mode, and temporal mode \cite{eisaman2011, reimer2019, senellart2017, aharonovich2016, wang2019}. However, efficiently collecting photons emitted by a QD with an objective lens or a single-mode fiber remains a challenge due to the high index contrast between the semiconductor surrounding the QD and air. Enhancement in the extraction efficiency can be achieved by incorporating nanophotonic structures including cavity coupled QD such as micropillar\cite{solomon2001single}, microdisk\cite{kiraz2001cavity}, photonic crystal cavity\cite{englund2005controlling}, and Bragg-grating bulls-eye microcavity\cite{ates2012bright}, waveguide coupled QD such as nanowire\cite{claudon2010highly} and photonic crystal waveguide\cite{lund2008experimental}, and other structures such as microlens\cite{gschrey2015highly} and optical antenna\cite{chen2018highly}. A collection efficiency of $\sim$80\% right after the objective lens is often demonstrated in the pillar-based and nanowire-based QD sources (Table 1 of \cite{senellart2017high}). In particular, QDs in tapered semiconductor nanowires (NW-QDs) have the additional benefit of nearly reflection-free emission into a mode propagating in free space and resembling a Gaussian-beam. This allows high photon-collection efficiency into a single-mode fiber as the nanowire taper can be engineered for mode-matching purposes. However, a common challenge for all these quantum dot platforms is achieving precise control of the frequency of emitted photons due to fabrication variations. This is particularly important when interfacing the QD photons with other quantum systems that can only interact with predefined photon frequencies, as is the case with atoms.
Implementing a hybrid quantum system that includes a QD-atomic interface, one needs to overcome the mismatch in the frequencies and linewidths of these two platforms. Atomic transitions at optical wavelengths are inherently well-defined by nature and have linewidths in the $\sim$MHz range. In contrast, the epitaxial growth of NW-QDs results in a standard deviation of approximately 5 nm ($\sim$6 meV) or more in the emission wavelengths and linewidths ranging from $\sim$GHz to $\sim$300 MHz for state-of-the-art NW-QDs \cite{laferriere2022unity, laferriere2023approaching}. While the linewidth issue can be addressed by using atomic ensembles with sufficiently large optical depths and with efforts to narrow the emission linewidth of the quantum dot \cite{reimer2016overcoming} or possibly by shaping the emitted photons \cite{sosnicki2023interface}, an in-situ, precise and broad-range tuning method for the quantum dot is necessary to properly match the frequency of the produced photons with the atomic transitions involved.

In recent years, various techniques have been reported that achieved tuning of the QD emission but many are not suitable for NW-QDs. The QD-emission tuning techniques include stress-induced wavelength shift by piezoelectric actuators \cite{sun2013strain, wildmann2015atomic, hofer2019tuning, cui2023coherent, kroh2019slow, trotta2016wavelength}, laser-induced crystallization of the surrounding environment \cite{grim2019scalable}, thermal \cite{leng2020tuning} and laser annealing \cite{dubowski2018precision}, local electric Stark field \cite{chu2023independent}, temperature-controlled tuning in a QD microlens \cite{bremer2020cesium}, surface acoustic wave (SAW) modulation  \cite{villa2017surface, tiwari2020radio, descamps2023dynamic}, laser-induced atomic intermixing in NW-QDs \cite{fiset2018situ}, stress induced by depositing dielectric material (SiO\textsubscript{2}) on the nanowire surface \cite{bouwes2012controlling}, and external magnetic fields applied to NW-QDs \cite{leandro2018nanowire}. 
However, these techniques are often platform specific, come with various degrees of limitations, and only few of them allow independent tuning of multiple quantum dots on the same chip. For example, piezoelectric actuators are widely used for planar structures but extremely challenging to employ in the NW-QDs due to their fragile nano/micro-sized vertical structures. 
The atomic intermixing in a nanowire is irreversible and offers blue shift only. Laser-irradiance-based tuning is non-linear, limiting the precision of tunability. Although deposition of dielectric materials on the nanowire can be reversible, achieving the necessary high-precision tunability for coupling to atoms is very challenging, as the wavelength shift cannot be monitored during the deposition process. The externally applied magnetic field tunes the QD emission by means of the Zeeman effect and diamagnetic shift, which is reversible, and precise to wavelength-match atomic transitions. However, this method requires large fields generated by superconducting coils, thus adding significantly to the experimental complexity, and it does not allow frequency control of multiple QDs independently. 
More importantly, the excitonic Zeeman spin splitting of the right ($\sigma^+$) and left ($\sigma^-$) circularly polarized photons degrades the entanglement fidelity proportionally to the amount of tuning, which poses a significant disadvantage when used for entangled-pair generation \cite{jovanov2012highly, hernandez2023effects}.


\begin{figure}[ht] 
\centering 
\includegraphics[width=0.95\textwidth]{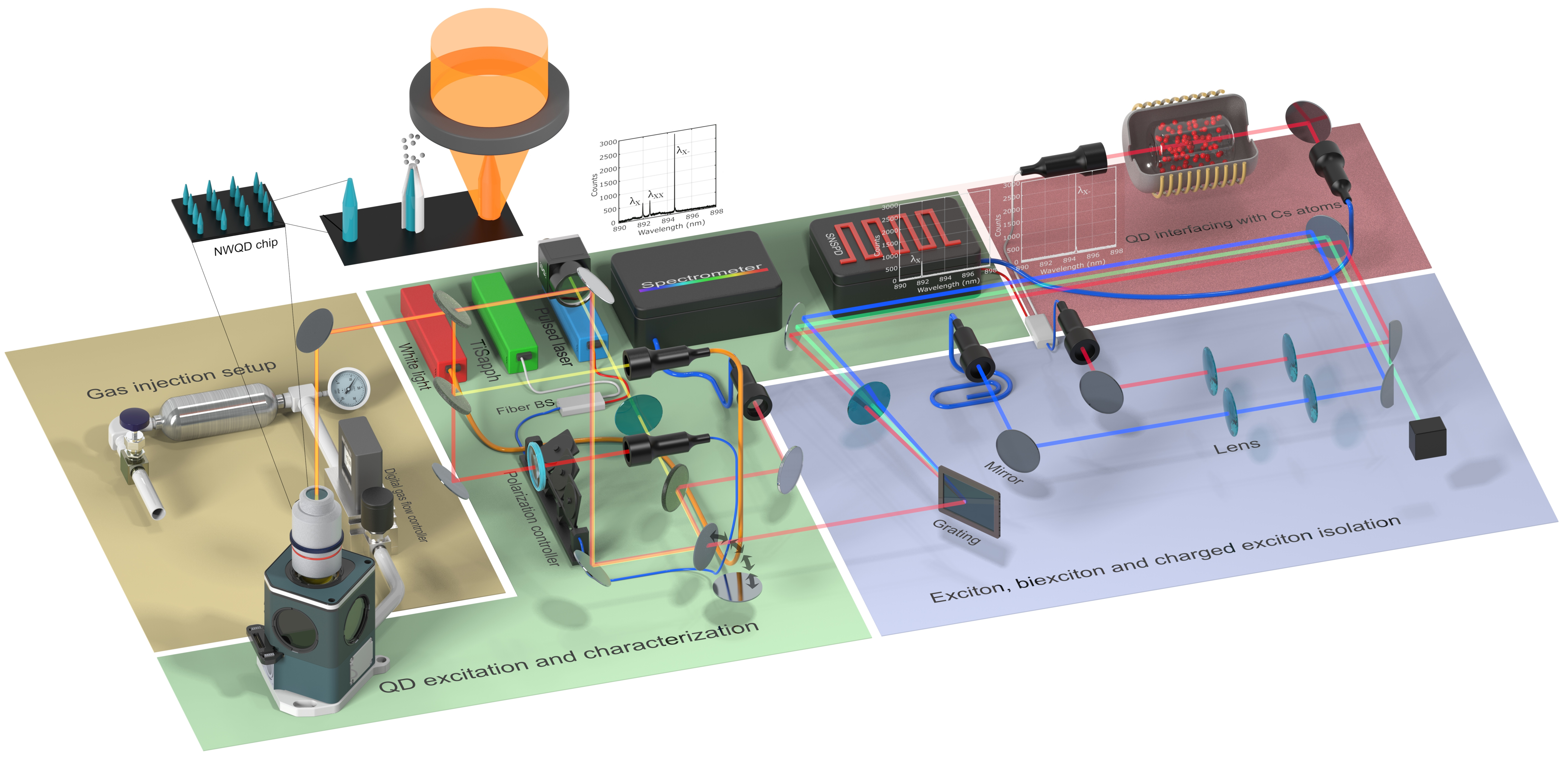} 
\caption{Experimental setup: NW-QD sample is kept at $\sim$4 K in a closed-cycle cryostat. A gas reservoir, digital mass flow controller, and a gas tube inside the cryostat are used for precision gas (N\textsubscript{2}) injection. CW Ti:Sapph or pulsed diode laser is used for QD excitation. The full QD emission spectrum can be seen on a spectrometer, or individual emission peaks ($X$, $XX$ and $X^-$) can be isolated using a transmission grating. In all the cases, QD photons are fiber coupled and used for different experiments.
}
\label{fig: exp setup}
\end{figure}

\section*{Results and discussion}\label{results}

Most of the properties of a QD, including the emission wavelength, depend on the size, shape and material of the QD. The QD emission wavelengths can be further controlled by applying some form of `stress' to the dot after it has been fabricated. In this work, we demonstrate an in-situ and reversible tuning of NW-QD emission by combining two processes: (i) the deposition of an inert gas (nitrogen) on the surface of the nanowire waveguide containing the QD, and (ii) the controlled evaporation of the deposited gas through local heating of individual nanowires using a focused laser beam. The first process is comparable to the ALD or PECVD deposition in the cleanroom \cite{bouwes2012controlling}, however, the method of deposition presented here is being implemented inside the cryostat without breaking the vacuum or cold cycle. Hence, a real time monitoring of the wavelength change is possible that is very critical to achieve high-precision tuning. 

\subsubsection*{Wavelength blue shift}
The NW-QDs  used in this work are InAsP quantum dots embedded in InP vertical nanowire  waveguides that have adiabatic tapering towards the top \cite{dalacu2021tailoring}. Room temperature gaseous nitrogen is injected into the cryostat (Fig. \ref{fig: exp setup}, inset), which then coats the surface of the chip with the NW-QDs, and gets solidified (nitrogen is solid below 63 K). The layer of solidified nitrogen puts a `compressive stress’ on the nanowire waveguide, which then strains the internal quantum dot structure. For intuitive understanding of the tuning mechanism, it is worth mentioning that the QD is theoretically a 0D material. Realistically, in this case it is a 3D disk shaped structure with approximately 20 nm diameter and 5 nm height embedded in a nanowire waveguide of 250 nm diameter at the base, 5 \textmu m in height and 1$^{\circ}$-2$^{\circ}$ adiabatic tapering at the top (sample growth technique and dimensions are in  Ref. \cite{laferriere2022unity}). The applied stress results in a slightly decreased size of the QD which in turns increases the energy gap between the valence band and the conduction band and a blue shift (decreased wavelength) of the QD emission occurs. One should note that while the technique of `gas deposition' for tuning photonic resonators has been demonstrated previously \cite{mosor2005scanning, dalacu2010deterministic, hausmann2013coupling}, the underlying physics involved in this approach is entirely different. When used to tune a micro- or nano-resonator,  the deposition of gas alters the effective refractive index of the resonator and consequently shifts its resonances. At the same time, when QDs are embedded in such resonators,  the gas deposition is not expected to affect their emission frequency. In this work, gas deposition occurs on the nanowire surface, which (i) may slightly alter the refractive index of the nanowire waveguide, yet has no impact on the emission frequency from the quantum dot, as it solely depends on the size and crystal composition (energy band diagram) of the quantum dot and its immediate surroundings; and (ii) imposes compressive stress on the nanowire and the internal quantum dot, leading to the emission frequency shift. The underlying physics is similar to that of references \cite{sun2013strain, bouwes2012controlling, rastelli2012controlling, fiset2018situ, seidl2006effect} and the stress is transferred to the embedded QD because of the small diameter of the semiconductor nanowire.

\begin{figure}[t]%
\centering
\includegraphics[width=0.95\textwidth]{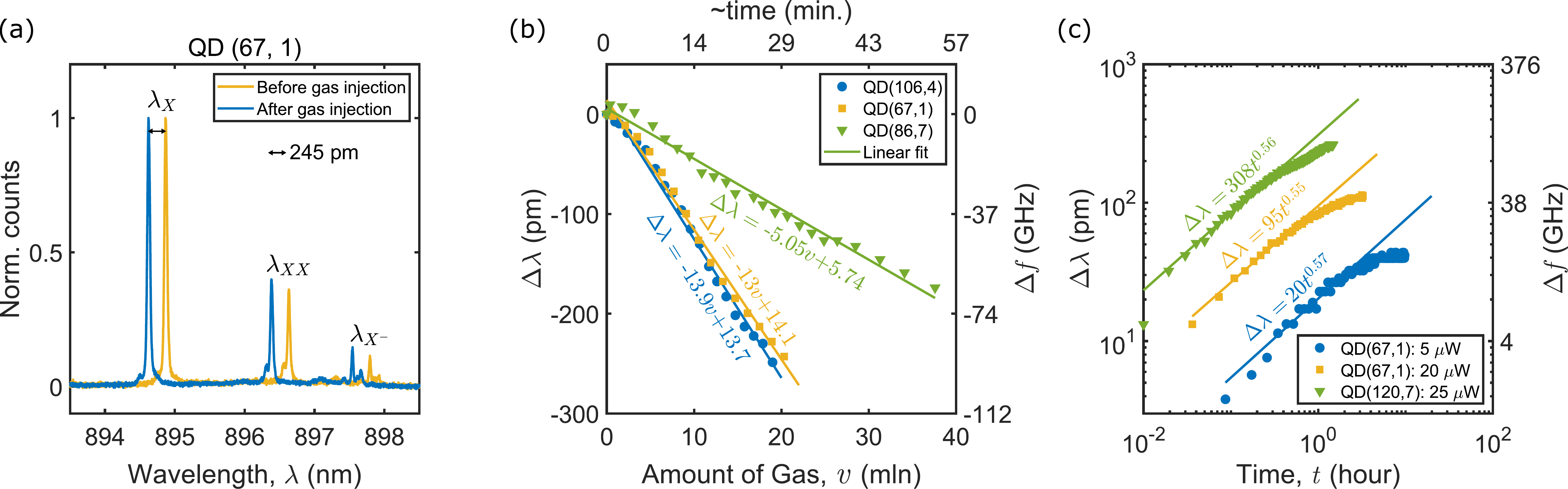}
\caption{(a) Quantum dot emission observed on the spectrometer before and after the gas injection. (b) Wavelength blue shift due to gas injection: measurements from three quantum dots (for example, QD(106, 4) means the quantum dot is located at 106\textsuperscript{th} row and 4\textsuperscript{th} column) are illustrated. The bottom (top) x-axis represents the amount of gas injected (time it required) and the left (right) y-axis corresponds to the change in emission spectrum in wavelength (frequency).  The circles, triangles and squares are measured data and the solid lines are linear fits. The corresponding fit equations are shown beside the lines. 
(c) Wavelength red shift due to laser ablation: the symbols are measured data and the solid lines are linear fits on a logarithmic scale ($\propto at^b$). Higher laser power yields faster ablation of N\textsubscript{2} ice, hence faster wavelength change initially. However, the shift of wavelength tends to saturate after some time as the amount of nitrogen evaporated and redeposited locally comes to an equilibrium.}
\label{fig: gas and laser summary plot}
\end{figure}

A diagram of our overall experimental setup is shown in Fig. \ref{fig: exp setup}. 
The blue shift of the emission spectrum as a result of the gas injection is illustrated in Fig. \ref{fig: gas and laser summary plot}(a), which shows the exciton (X), biexciton (XX), and negatively charged exciton (X$^-$) emission peaks of one of our QDs. The amount of change for each of the QD emission peaks is the same ($\sim$245 pm). The change of the wavelength is proportional to the amount of gas injected as presented in Fig. \ref{fig: gas and laser summary plot}(b). The rate of change for different QDs differ partly due to the slight variation of the internal crystalline structure of the individual nanowire waveguides and of the quantum dots inside them. Additionally, each nanowire has a slightly different distance from the gas injection outlet, which affects the amount of deposited gas. Approximately 100 GHz of tuning is comfortably attainable in less than an hour. To achieve a target emission wavelength, the emission spectrum is continuously monitored on a spectrometer and the gas flow rate can be changed or stopped as required. The maximum shift observed is $\sim$832 pm or $\sim$313 GHz or $\sim$1.29 meV (see supplementary, Fig. \ref{fig: wavelength change in QDs}). However, this is not a fundamental limit and a larger tuning range can likely be achieved with a more sophisticated gas injection setup (see Methods).

\subsubsection*{Wavelength red shift}

The blue shift in the NW-QD emission can be reversed by removing the solid nitrogen deposited on the nanowire surface. However, an incremental increase of the temperature of the whole NW-QD sample near the operating point ($\sim$4 K) through thermal heating  would leave the deposited N\textsubscript{2} intact as it has a melting point of $\sim$63 K. On the other hand, heating the sample past the melting point would remove the deposited nitrogen completely and the emission wavelength would return to the initial state. Simple heating thus can't provide precision tuning but it is convenient for complete removal of nitrogen from the whole QD sample. Laser ablation on the other hand \cite{niino2002laser}, can be applied locally on an individual NW-QD and in a very precise fashion by controlling the laser power, wavelength, exposure time, and the beam size. As a result, laser ablation can also be used to tune multiple NW-QDs on the same chip independently. We use focused CW Ti:Sapph laser at a wavelength of 910 nm with varying intensity and exposure time for laser ablation of the deposited nitrogen.  The experimental results are presented in Fig. \ref{fig: gas and laser summary plot}(c). Higher intensity or longer exposure proportionally red-shifts the emission wavelength initially, however, saturates after a certain exposure time. This is likely due to the illuminated spot with the nanowire reaching an equilibrium of evaporation and re-deposition of N\textsubscript{2} ice and gas, respectively. Similar to the gas deposition, laser exposure could also tune the QD emission frequency by more than 100 GHz in hour(s). By combining the blue and red shift achieved with the gas deposition and the laser ablation, very precise tuning of the quantum dot emission necessary to match the atomic transitions is attainable. It is worth mentioning that the initial emission wavelength has to be above the desired wavelength as this method allows blue-tuning followed by red-tuning, but not the other way around.

\subsubsection*{Single-photon properties}

To assess the impact of deposited nitrogen on the properties of quantum dot emission, we performed several quantitative measurements before and after the wavelength was tuned via gas deposition. Specifically, the second order auto-correlation function, lifetime, linewidth and fine-structure splitting (FSS) were measured for the exciton emission of QD(67, 1).

\begin{figure}[b]%
\centering
\includegraphics[width=0.85\textwidth]{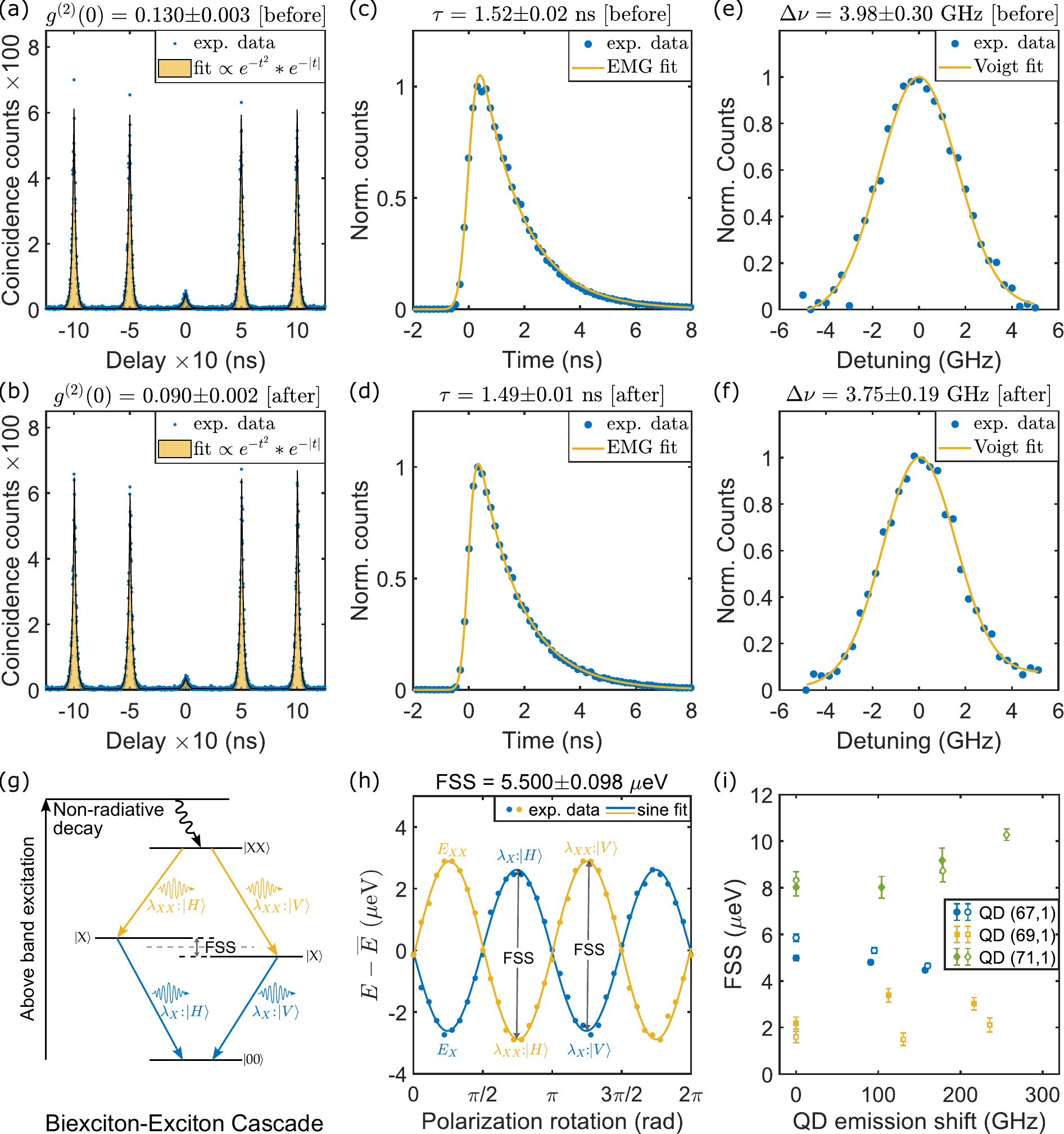}
\caption{Second order auto-correlation ($g^{(2)}(\tau)$) (a-b), lifetime ($\tau$) (c-d), linewidth ($\Delta \nu$) (e-f) and fine-structure splitting (FSS) (g-i) measurements of exciton emission of QD(67, 1) before and after the wavelength tuning (460 pm or 172 GHz for (a-f)) by gas injection. Details of the fitting are in the method section. For $g^{(2)}(0)$, $\tau$ and $\Delta \nu$, the values are similar before and after the wavelength tuning which suggests that these emission properties are negligibly affected by the deposited nitrogen. 
(g) Biexciton-exciton cascade: QD emits either $\vert H\rangle$ or $\vert V\rangle$ polarized photons with equal probabilities. FSS = $E_{\lambda_{XX:\vert V\rangle}} - E_{\lambda_{X:\vert V\rangle}}$ = $E_{\lambda_{X:\vert H\rangle}} - E_{\lambda_{XX:\vert H\rangle}}$. 
(h) FSS measurement: Evolution of the change of exciton and biexciton photon energies as a function of polarization rotation. The measured FSS is (5.500$\pm$0.098) $\mu$eV or (1.330$\pm$0.024) GHz.
(i) FSS measurement for three QDs at different tuning and for two measurement cycles, warm-up and cool-down in between. The solid and open symbols are for first and second measurement cycle, respectively. Negligible change is observed, either favorably or adversely, for up to $\sim$100 GHz, however, limited to $<$2 $\mu$eV of change of their respective starting values for tuning in the range of $\sim$100-300 GHz.
}
\label{fig:g2_lifetime_linewidth}
\end{figure}

In Fig. \ref{fig:g2_lifetime_linewidth}(a-b), we present the results of performing the second order auto-correlation measurement using a Hanbury Brown and Twiss setup. The measured $g^{(2)}(\tau)$ is less than 0.5, indicating a single-photon source. The measurement is performed at $\sim$10\% of the saturation counts, and the $g^{(2)}(0)$ values before and after the wavelength shift are ($0.130 \pm 0.003$) and \mbox{($0.090 \pm 0.002$)}, respectively (see Methods).
The lifetime of the QD excitonic state $(\tau)$ is measured as the delay between the moment of absorbing a photon from the excitation laser and the moment of emitting an exciton photon. The measurement is similar to the $g^{(2)}(\tau)$ measurement except that the start signal is the clock pulse. The results are presented in Fig. \ref{fig:g2_lifetime_linewidth}(c-d). The exciton lifetimes of the measured quantum dot are ($1.52\pm 0.02$) ns and ($1.49\pm 0.01$) ns before and after the wavelength tuning, respectively (see Methods). 
The linewidth $(\Delta\nu)$ is defined as the FWHM of the power spectral density of the emission spectrum. The measurement is performed by sending the exciton emissions through a scanning Fabry-P\'erot interferometer, and fitting the experimental data with a Voigt profile (see Methods). The measured linewidths before and after the frequency tuning are ($3.98 \pm 0.30$) GHz and ($3.75 \pm 0.19$) GHz, respectively, which are within the measurement error limit (Fig. \ref{fig:g2_lifetime_linewidth}(e-f)).

The energy band diagram of a NW-QD illustrated in Fig. \ref{fig:g2_lifetime_linewidth}(g) shows that the electron-hole pairs can emit biexciton-exciton photons through either left ($\vert H\rangle$ polarization) or right ($\vert V\rangle$ polarization) decay paths. The energy difference between the middle fine-structure energy bands (FSS) is zero for a uniform and unstrained QD, but fabrication imperfections lead to a small value, typically a few $\mu$eVs for our QDs. This affects entanglement fidelity if the NW-QD is used as a source of entangled photon pairs \cite{fognini2019dephasing}. Wavelength tuning is achieved by straining the QD with deposited nitrogen ice, prompting the measurement of FSS changes arising from the  process. The experimental result is illustrated in Fig. \ref{fig:g2_lifetime_linewidth}(h), where the measured FSS is (5.500±0.098) μeV (see Methods). Figure \ref{fig:g2_lifetime_linewidth}(i) depicts FSS changes in three QDs with different tuning. We observe a negligible change for up to $\sim$100 GHz of tuning, and a change of up to 2 $\mu$eV for $\sim$100-300 GHz of tuning.

Overall, the gas deposition does not seem to significantly change the measured properties of the NW-QD photon source, at least as far as we were able to observe and with the exception of the FSS. 
Moreover, the tuning is stable on the order of atomic resonances ($\pm$ 1 GHz) for days (see supplementary, Fig. \ref{fig: Supplemantary stability}).

\subsubsection*{Single-photons tuned to atomic transitions}

\begin{figure}[t]%
\centering
\includegraphics[width=0.9\textwidth]{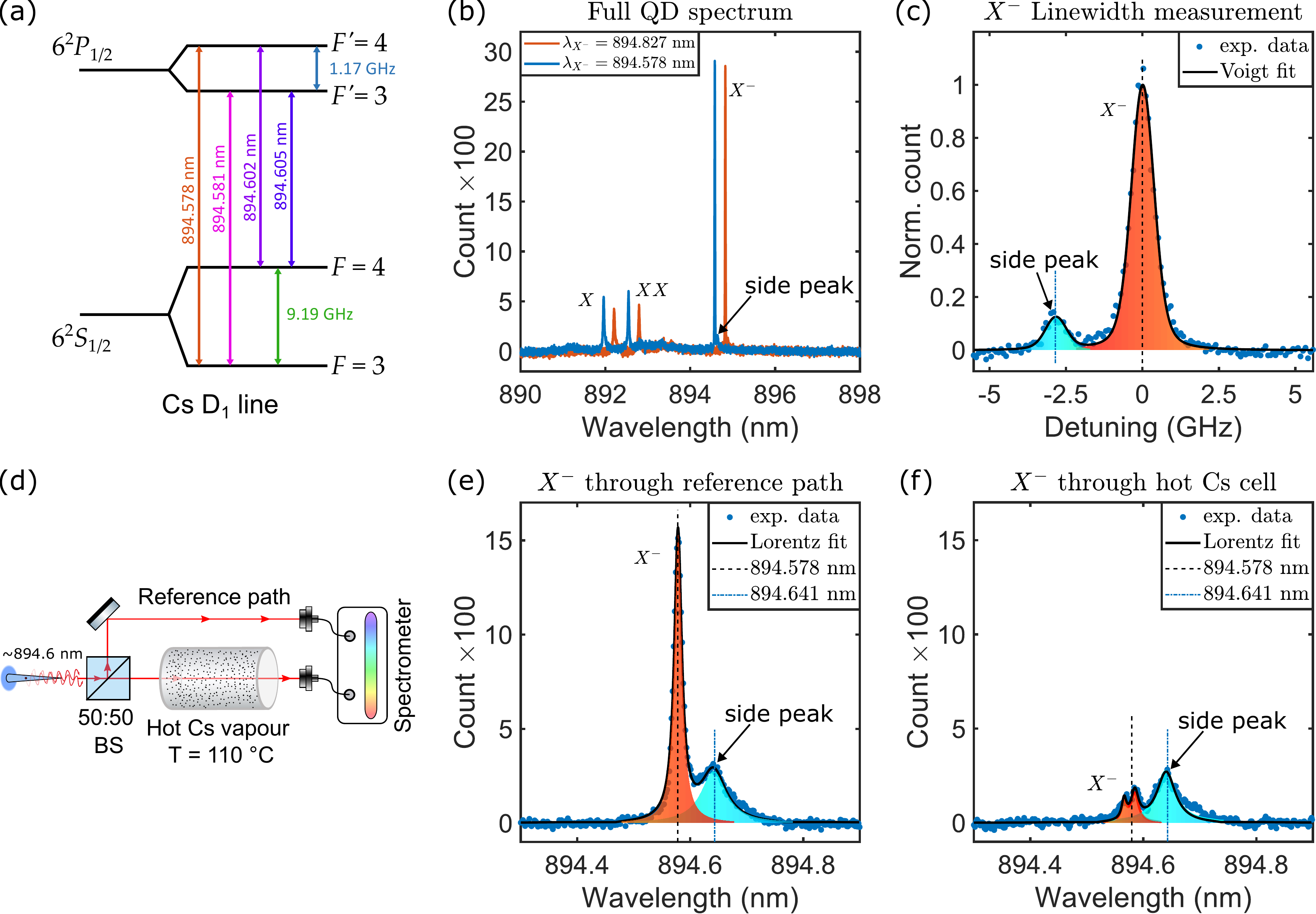}
\caption{(a) Cs D\textsubscript{1} transition hyperfine structure, with energy splittings between the hyperfine energy levels. 
(b) QD(43, 7) emission spectrum before (red) and after (blue) the tuning. The negatively charged exciton ($X^-$) is tuned at $F=3 \rightarrow F^\prime=4$ transition. A tiny side (parasitic) peak is visible near the $X^-$ emission. 
(c) Linewidth measurement using a scanning Fabry-P\'erot cavity shows the presence of the side lobe. $\text{FWHM}_{X^{-}} = (900\pm24)$ MHz and $\text{FWHM}_{side\, peak} = (830\pm192)$ MHz.
(d) Schematic of the experimental setup. 
(e-f) The spectrum of the reference path shows transmitted photons at 894.578 nm wavelength while the $X^-$ photons going through the hot Cs cell are mostly absorbed. The side peak (at 894.641 nm) is unaltered due to its separation of \mbox{$\sim$36 pm} or $\sim$14 GHz from the nearest $F=4 \rightarrow F^\prime=3$ transition.
}
\label{fig: side_peak}
\end{figure}

Next, we proceed to demonstrate linking of the NW-QD with atomic ensemble through resonant absorption and slow single-photons in hot caesium vapour as the quantum dot emission is tuned across the Cs D\textsubscript{1}-line. 
While the interfacing of quantum dots with atomic ensembles has been reported in the recent past \cite{akopian2011hybrid, kroh2019slow, trotta2016wavelength, bremer2020cesium, cui2023coherent}, the purpose of this demonstration is to showcase the precision of our tuning method.
As shown in  Fig. \ref{fig: side_peak}(a), caesium D\textsubscript{1}-line has four hyperfine transitions, two at the ground state ($F$ = 3, 4 at $6^2S_{1/2}$) and two at the excited state ($F^\prime$ = 3, 4 at $6^2P_{1/2}$). These transitions are Doppler broadened to $\sim$GHz linewidths above 100 $^{\circ}$C. The ground and excited states have energy gaps of $\sim$9.2 GHz and $\sim$1.2 GHz, respectively.

To efficiently interface the quantum dot photons with the D\textsubscript{1}-line, quantum dot emission with a narrow linewidth -- preferably sub-GHz -- near the atomic resonances is necessary. In our sample, we observe a typical linewidth of the exciton and biexciton emission of $\sim$3.5-4 GHz, while the charged-exciton has a sub-GHz linewidth (Supplementary Fig. \ref{fig: Supplemantary linewidth}). We select a quantum dot, \mbox{QD(43, 7)}, with the $X^-$ emission peak originally slightly above the $F=4 \rightarrow F^\prime=3$ transition wavelength and use the gas deposition to reduce (increase) the emission wavelength (frequency) and scan through the D\textsubscript{1}-line. Similarly, local heating by focused laser exposure is used to scan in the opposite direction. The full emission spectrum of QD(43, 7) -- original and after gas tuned to one of the resonances ($F=3 \rightarrow F^\prime=4$) -- is depicted in Fig. \ref{fig: side_peak}(b). A tiny `parasitic' emission peak near the $X^-$ is observed on the spectrometer, which indicates that the emission of this NW-QD occasionally jumps frequency and produces photons of slightly different wavelength. This `parasitic' emission is also visible in the cavity scan used in the linewidth measurement (Fig. \ref{fig: side_peak}(c)). To investigate the interaction of these two emission peaks with Cs atoms, resonant $X^-$ photons are sent through a 50:50 beam splitter; one path is directed through a hot caesium glass cell at 110 $^{\circ}$C to the spectrometer while the other path is going to the spectrometer directly (Fig. \ref{fig: side_peak}(d)). The photons going directly to the spectrometer show the $X^-$ peak at 894.578 nm wavelength and a minor side peak at 894.641 nm wavelength (Fig. \ref{fig: side_peak}(e)). However, the photons going through the hot Cs cell are mostly absorbed at 894.578 nm while the side peak is unaltered (Fig. \ref{fig: side_peak}(f)). This is due to its separation of $\sim$36 pm or $\sim$14 GHz from the nearest $F=4 \rightarrow F^\prime=3$ transition.

\begin{figure}[b]%
\centering
\includegraphics[width=0.9\textwidth]{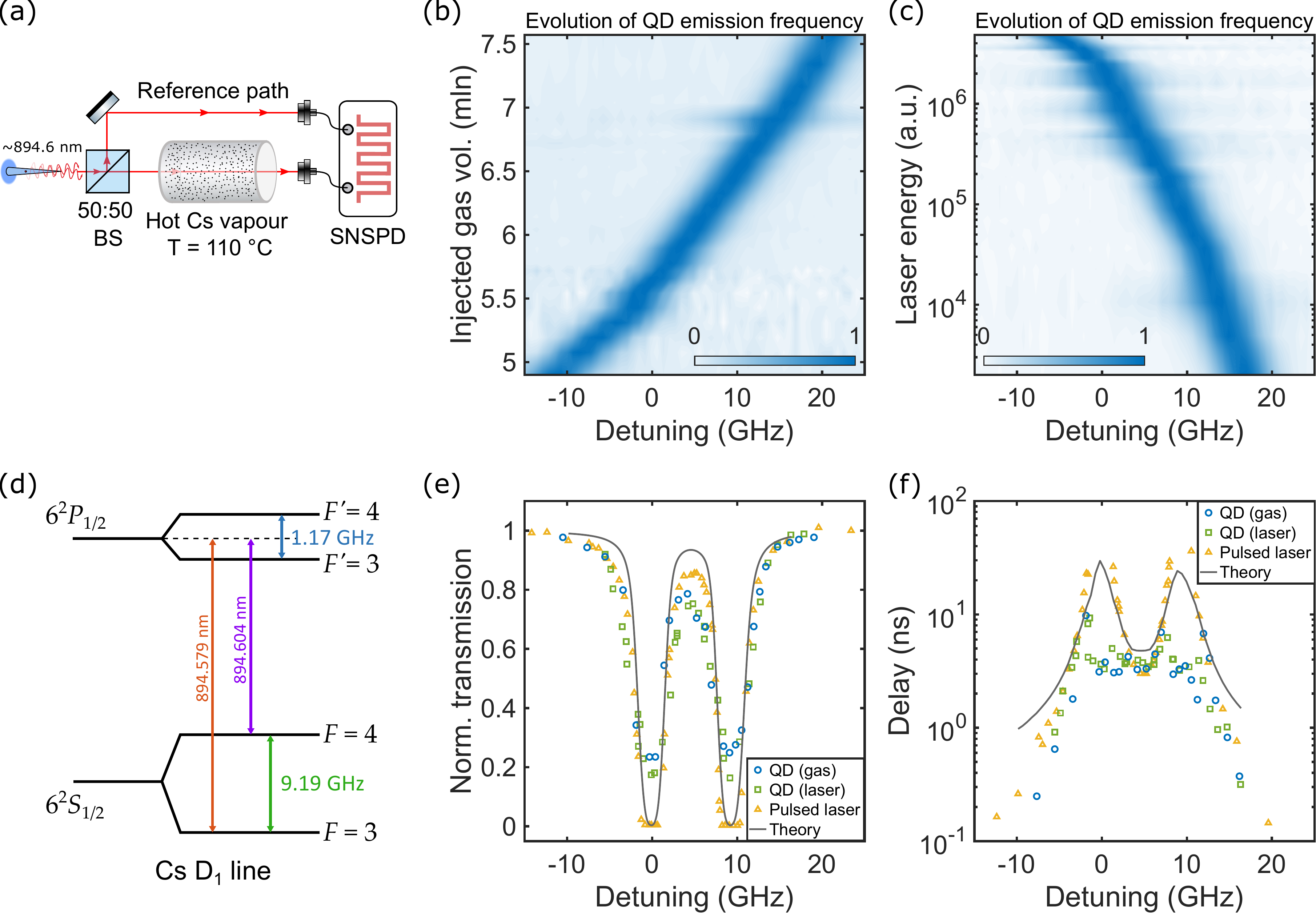}
\caption{ Resonant absorption in hot Cs vapour: (a) Schematic of the experimental setup. 
(b-c) Evolution of QD emission frequency (negatively charged exciton, $X^-$, linewidth $\sim$900 MHz) due to gas deposition (lower to higher frequency) and heating laser exposure (higher to lower frequency). The FWHM of the depicted spectrum is $\sim$5 GHz which is resolution-limited by the spectrometer and not the true linewidth of the QD.
(d) Cs D\textsubscript{1} transition hyperfine structure. These transitions are Doppler broadened to $\sim$GHz linewidths above 100 $^{\circ}$C. Hence, for this experiment we assume a single energy level for the excited state at the original location of $6^2P_{1/2}$ due to its $\sim$GHz energy gap.
(e) Normalized transmission vs detuning: scanning of QD is done using gas injection (blue circles) and heating laser exposure (green squares). In both cases, the resonant transmission only drops to $\sim$0.2, mainly due to the linewidth of our NW-QD and the presence of the side peak in the NW-QD emission spectrum. In contrast, the laser pulse linewidth is $\sim$220 MHz 
which results in near zero transmission when the pulse is resonant with Cs transitions. The experimental results match well with our theory and numerical simulations.
(f) Delay vs detuning: delayed QD photons on both sides of the resonances and in between follows the theory simulation, however, is tricky to measure near the resonances due to multiple peaks of the transmitted pulses, especially for the wider pulses.
}
\label{fig: scanning of QD}
\end{figure}

To measure the transmission change of $X^-$ photons across the caesium D\textsubscript{1}-line, the QD emission is continuously tuned either by using the gas deposition or by using the heating laser exposure. The evolution of the QD emission frequency as a function of injected gas and laser energy is depicted in Fig. \ref{fig: scanning of QD}(b, c). It is worth mentioning that the resolution of the scanning depicted here using gas deposition (heating laser exposure) is ($1.18\pm0.57$) GHz (($0.73\pm0.44$) GHz) but these could be further reduced if needed. The QD emission is tuned to a particular frequency and half of the photons are sent directly to a SNSPD while the other half is directed through a hot Cs cell before going into the SNSPD (Fig. \ref{fig: scanning of QD}(a)). The same experiment is repeated with a pulsed laser for reference. Fig. \ref{fig: scanning of QD}(e) shows the experimental result, where the normalized transmission is defined as the ratio of the total photons between the caesium cell path and the reference path and zero detuning is defined at the $F=4 \rightarrow 6^2P_{1/2}$ transition (Fig. \ref{fig: scanning of QD}(d)). 
The linewidth of the QD photons is ($900\pm24$) MHz while the linewidth of the pulsed laser is $\sim$220 MHz. This also allows about four-times better resolution for the laser scanning compared to the tuning of the QD-emitted photons. The solid black curve is the simulation result from theory (details in Methods and in Supplementary). For photons emitted by the NW-QD as it is tuned, the resonant transmission drops only as low as $\sim$0.2 instead of zero mainly due to (i) the linewidth of the photons emitted by our NW-QD 
and (ii) the presence of the nearby parasitic emission peak (see Supplementary Fig. \ref{fig: Cs_cell_transmission_method} for details). In contrast, the laser pulse is completely absorbed near the resonances due to its narrower linewidth and cleaner spectrum with no side peak. The photon delay through the hot cell during the scanning is depicted in Fig. \ref{fig: scanning of QD}(f). Photon delays on both sides of the resonances and in between follow the predictions obtained from our theoretical calculations. However, it is difficult to determine delays near the resonances as the transmitted pulses display multiple peaks. This becomes especially pronounced for the QD-emission pulses which have larger bandwidth than the laser pulses. Hence, to further study the effect of warm caesium atoms on the group velocity of QD-emitted photons near the atomic resonances, the following experiment is performed. 


\begin{figure}[b]%
\centering
\includegraphics[width=1\textwidth]{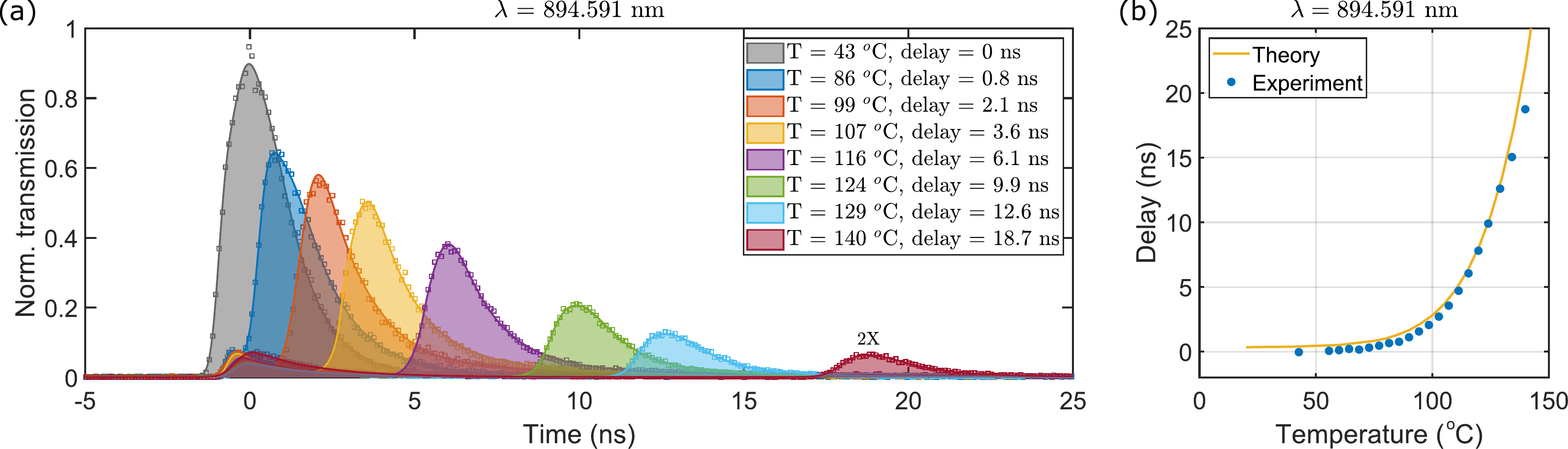}
\caption{Slow light in hot Cs vapour: (a) single-photon pulse propagation through Cs vapour at different temperatures. Transmission reduces with the increase of temperature due to increasing optical depth. Maximum delay of $\sim$18.7 ns is observed at 140 \textsuperscript{o}C. 2X multiplier at T = 140 \textsuperscript{o}C is for better visualization.
(b) single-photon delay vs Cs vapour temperature and theory simulation.
}
\label{fig: delay vs temp}
\end{figure}

The QD emission is tuned and parked at the mid-point between the \mbox{$F=3 \rightarrow 6^2P_{1/2}$} and $F=4 \rightarrow 6^2P_{1/2}$ hyperfine transitions at 894.591 nm. The temperature of the caesium cell is varied from room temperature to \mbox{140 $^{\circ}$C} and the normalized transmission through the cell is continuously monitored. As the temperature increases, the photon pulses going through the cell get delayed, absorbed and broadened compared to the reference path (Fig. \ref{fig: delay vs temp}(a)). This is due to the increase of both the real ($n$) and the imaginary part ($\kappa$) of the complex refractive index of the atomic ensemble as the density of the caesium vapour increases with temperature. 
 A maximum delay of 18.7 ns or a 75-fold decrease in the group velocity of photons is observed for the $X^-$ emissions at the highest temperature, along with a very low transmission. A plot of delay vs. temperature is also depicted in Fig. \ref{fig: delay vs temp}(b). The experimental data fit very well with the theory simulation. It is worth noting that the temperature change has no effect on the location of the pulse component resulting from the `parasitic' $X_{side peak}$ emission, which stays near 0 ns. As mentioned in the previous section, this side peak is $\sim$63 pm detuned from the $X^-$ peak; i.e. the location of the side peak is at 894.654 nm in this case. 

\section*{Conclusion}\label{sec13}

In conclusion, we have demonstrated a full bi-directional control over the frequency of photons emitted by a nanowire quantum dot. This has allowed us to address specific optical transitions in caesium atoms with the QD-emitted photons. Additionally, our approach allows independent tuning of individual quantum dots on the same chip for a type of QD that is not compatible with tuning using piezo-electric actuators. Moreover,  the tuning is achieved inside the cryostat without breaking the vacuum or cold cycle and without applying large magnetic fields. We demonstrate reversible tuning exceeding 300 GHz with sub-GHz precision. The tuning process has no noticeable effect on many key emission properties of NW-QDs, such as $g^{(2)}(0)$, lifetime, linewidth, and photon extraction efficiency. 
Furthermore, there is either no or negligible observable effect on the fine structure splitting (FSS) for tuning up to 100 GHz, which is crucial for maintaining entanglement fidelity. The tuning process is reasonably fast, taking only minutes, and the resulting emission wavelength remains stable for time scales of days. 
The demonstrated control over the emission frequency of the NW-QD with a relatively simple hardware opens several new and exciting research and application avenues. Direct interfacing of the bright single-photon source based on a QD with atomic platforms makes it possible to move from experimental demonstrations with `coherent pulses at single-photon level' to probing and control of atomic systems with true single-photon pulses.  We thus expect our work to lead to significant improvements in the performance of quantum repeaters based on both heralded and unheralded atomic quantum memories, as well as to enable near-unity frequency conversion of single-photons using atomic ensembles.
Additionally, creating collective states in atomic ensembles with true single-photon pulses arriving with high repetition rate has the potential to facilitate practical generation of new non-classical states of light and enhance the speed of single-photon transistors based on atomic platforms, as demonstrated in Ref. \cite{chen2013all}. Lastly, the emission tuning via gas deposition can possibly be extended to other single-photon sources based on solid-state quantum emitters embedded in thin waveguides, such as lattice defects in diamond nanowires \cite {hausmann2010fabrication} or rare earth ions embedded in nanobeams \cite{zhong2015nanophotonic}.

\newpage
\backmatter

\section*{Methods} $\,$

 \textbf{Limit of gas deposition tuning:}
 With the current gas injection setup, excessive amount of gas injected into the cryostat results in condensation forming on the cryostat windows, which makes them partially opaque. This reduces the visibility of NW-QD emission, which makes the optical alignment harder and reduces the collection efficiency. A tuning range larger than the $\sim$313 GHz reported here can likely be achieved with a more targeted gas injection setup or using a photon collection setup that does rely on optical access to the NW-QD through a window.
 
\textbf{Laser ablation of deposited gas:}
The specific wavelength of 910 nm is chosen because: (i) the laser wavelength is $>$10 nm below the bandgap of the QD so that the ablation laser doesn't create electron-hole pairs and hence doesn't emit QD photons, (ii) the tuning range of the CW Ti:Sapph laser used here is 700 nm to 1000 nm and the emission power of the laser peaks at 770nm and decreases with increasing wavelength, especially past 900nm, and (iii) the ablation laser wavelength is needed to be outside of the spectrometer window ($890\, \text{nm} > \lambda > 900\, \text{nm}$) for continuous monitoring of the quantum dot emission under pulsed above-band excitation at 823 nm.

\textbf{QD spectroscopy:}
The QD excitation is done with either a CW Ti:Sapph laser or a pulsed diode laser at 823 nm for above-band excitation with a pulse width of $\sim$80 ps and a repetition rate of 20 MHz to 80 MHz depending on the measurement. A white light source and a camera is used to locate the desired NW-QD. The emission from the quantum dot is either fiber coupled to a spectrometer or sent through a transmission grating before coupling to a fiber. The former is useful to look at the full emission spectrum including exciton ($X$), biexciton ($XX$) and charged-exciton ($X^-$) photons while the later is necessary to isolate the individual emission peaks. Single Photon Counting Modules (SPCMs) and/or Superconducting Nanowire Single Photon Detectors (SNSPDs) are used for single-photon detection and measurements, depending on the experiment. The collection efficiency at the first lens (objective lens) is approximately 40\%, considering only photons emitted in the upward direction. 

\textbf{$g^{(2)}(0)$ measurement:} The measurement is defined as: 
\begin{linenomath} 
\[g^{(2)}(0)=\frac{A_0}{\sum_{i=1}^{N} A_i}\pm \delta\]
\end{linenomath} 
Here, $A_0, A_i=$ area under the center peak and side peak, respectively and $\delta=$ error. The area is calculated as the sum of experimental counts over the fitted FWHM spread. The fitting for the $g^{(2)}(0)$ measurement is a convolution of a Gaussian pulse (instrument response) and a double sided exponential function (quantum dot decay).  The $g^{(2)}(0)$ is averaged over 12 side-peaks (only 4 is shown in Fig. \ref{fig:g2_lifetime_linewidth}(a-b)) with 156.25 ps time bin resolution and the standard deviation is represented as the error.

\textbf{Lifetime measurement:} The fitting for the lifetime measurement is the exponentially modified Gaussian (EMG) function which is a convolution of a Gaussian pulse (instrument response) with an exponential function (quantum dot decay). EMG function is defined as: 
\begin{linenomath} 
\[f(t)=\frac{1}{2}e^{-\tau(t-\frac{\sigma^2\tau)}{2}}[1+\text{erf}\,(\frac{t-\sigma^2\tau}{\sqrt{2}\sigma})]\]
\end{linenomath} 
Here, $\tau$ is the exponential decay constant (defined as the lifetime of the quantum dot) and $\sigma$ is the Gaussian width. The error is defined as the 95\% confidence interval of the fitting parameter $\tau$.

\textbf{Linewidth measurement:} The scanning of the Fabry-P\'erot cavity (10 GHz FSR, 67 MHz resolution) is done using a periodic triangular voltage from 0 to 10 V\textsubscript{P-P} in 1,000 s which is equivalent to a $\sim$20 GHz sweep. The output of the cavity is detected using a single-photon detector and the experimental data is fitted with a Voigt profile, defined as the convolution of a Gaussian profile (inhomogeneous broadening and instrument response) with a Lorentzian profile (homogeneous broadening):
\begin{linenomath} 
\begin{gather*}
     V(x;\sigma, \gamma) \equiv \int_{-\infty}^{\infty} G(x';\sigma) L(x-x';\gamma) dx'\\
     G(x;\sigma) \equiv \frac{e^{-x^2/(2\sigma^2)}}{\sigma \sqrt{2\pi}}; \qquad 
     L(x;\gamma) \equiv \frac{\gamma}{\pi(x^2+\gamma^2)}
\end{gather*}
\end{linenomath} 
Here, $x$ is the shift from the center, $G(x;\sigma)$ and $L(x;\gamma)$ are the centered Gaussian and Lorentzian profile, respectively. The integral is evaluated as:

\begin{linenomath} 
\begin{gather*}
     V(x;\sigma, \gamma) = \frac{\text{Re}[w(z)]}{\sigma \sqrt{2\pi}} \\
     w(z) = e^{-z^2} \left(1 + \frac{2i}{\sqrt{\pi}}\int_{0}^{z} e^{t^2} \text{d}t \right); \qquad
     z = \frac{x+i\gamma}{\sigma \sqrt{2}}
\end{gather*}
\end{linenomath} 

The FWHM of the Voigt profile is the linewidth of the quantum dot emission, defined as $f_V \approx 0.5346 f_L+\sqrt{0.2166f_L^2+f_G^2}$. Here, FWHM of the Gaussian $(f_G)$ and Lorentzian $(f_L)$ profiles are defined as $f_G = 2\sigma\sqrt{2\ln(2)}$ and $f_L = 2\gamma$, respectively.

\textbf{Fine-structure splitting (FSS) measurement:} The measurement is performed by sending the QD emission through a rotating half-wave plate (HWP) and a fixed linear polarizer. Depending on the HWP position either $\vert H\rangle$ or $\vert V\rangle$ or a mixture of both polarized photons are collected on the spectrometer. The change in the peak emission energy as a function of polarization rotation (2$\times$HWP rotation) is plotted. The experimental data is then fitted with a sine function. The FSS is measured as the difference of exciton and biexciton photon energies when only the vertical or horizontal photons are selected.

\textbf{Caesium spectroscopy:}
A 7.5 cm long glass spectroscopy cell with caesium inside is used for experiments involved in interfacing QD photons with caesium atoms. The cell is kept inside a metal box with a ribbon heater wrapped around it.

\textbf{Theory simulation:} In order to estimate features of pulse propagation, the atomic vapour's polarizability is considered as a medium with four Lorentzian resonances (namely, $6S_{1/2}$ F = 3 $\leftrightarrow$ $6P_{1/2}$ F$^\prime$ = 3, $6S_{1/2}$ F = 3 $\leftrightarrow$ $6P_{1/2}$ F$^\prime$ = 4, $6S_{1/2}$ F = 4 $\leftrightarrow$ $6P_{1/2}$ F$^\prime$ = 3, $6S_{1/2}$ F = 4 $\leftrightarrow$ $6P_{1/2}$ F$^\prime$ = 4). Temperature affects the vapour's density and Doppler broadens the transitions. Output pulses are computed under the slowly-varying envelope approximation, where the input pulse is set as a Gaussian with a bandwidth matching the measured value. This propagation lends both, the transmission and delay, which are plotted in Fig.\ref{fig: scanning of QD}(e) and (f). 

 When the pulse's central frequency is parked midway between the two hyperfine states, the two excited states ($6P_{1/2}$ F$^\prime$ = 3, $6 P_{1/2}$ F$^\prime$ = 4) is approximated as a single effective transition. Pulse delay is shown to occur due to the dispersive region between the two Lorentzian resonances ($6S_{1/2}$ F = 3 $\leftrightarrow$ $6P_{1/2}$ and $6S_{1/2}$ F = 4 $\leftrightarrow$ $6P_{1/2}$). This delay as a function of temperature of the vapour is plotted in Fig. \ref{fig: delay vs temp}(b).

\section*{Acknowledgments}
This research was undertaken thanks in part to funding from the Canada First Research Excellence Fund. It was also supported by Canada Foundation for Innovation, Industry Canada, NSERC’s Discovery grant and Research Tool and Instruments grant, and by Ontario’s Ministry of Innovation Early Researcher Award. A special thanks to Shahriar Ashad Durjoy for the 3-dimensional model of the experimental setup in Fig. 1.

\section*{Author contributions}
M.B., R.A.M., and M.E.R. conceived the idea. R.A.M. designed and built most of the experimental setup, carried out the measurements and analyzed the data. S.V. carried out the numerical simulations. S.V., D.B., P.A., J.Q., Y.Y., M.Z., B.S., and K.R. helped to build parts of the experimental setup. P.J.P. and D.D. fabricated the nanowire quantum dot chip and D.B. and J.Q. carried out the initial measurements to characterize the quantum dots. R.A.M. wrote the manuscript with the contribution from all authors. M.B. and M.E.R. supervised the project.

\bmhead{Competing interests}
The authors declare no competing interests.

\bibliography{2bibliography}
\newpage

\setcounter{section}{0}
\setcounter{figure}{0}
\makeatletter
\renewcommand \thesection{S\@arabic\c@section.}
\renewcommand\thetable{S\@arabic\c@table}
\renewcommand \thefigure{S\@arabic\c@figure}
\renewcommand{\citenumfont}[1]{S#1}
\makeatother

\begin{center}
    \large \textbf{Supplementary information}
\end{center}

\section{Wavelength change variations in different QDs}

\begin{figure}[ht]%
\centering
\includegraphics[width=1\textwidth]{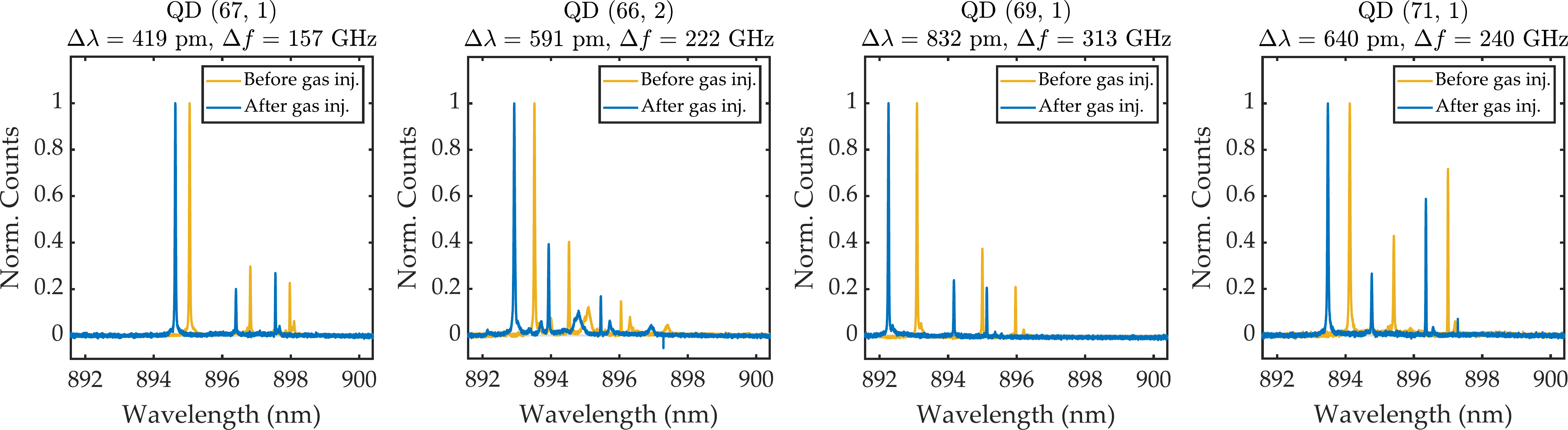}
\caption{Variation in change of wavelengths for different QDs: The measurement is done by sending $\sim$20 mln nitrogen gas into the cryostat. During the measurement, four different QDs at different coordinates are monitored. A minimum change of 157 GHz is observed for QD(67, 1) and a maximum change of 313 GHz is observed for QD(69, 1). The difference of twice the amount of change between minimum and maximum shift is partly due the internal crystalline structure of the nanowire quantum dot and partly due to the relative location of the NW-QD with respect to the gas injection orifice.
}
\label{fig: wavelength change in QDs}
\end{figure}

\section{Stability of the emission after the tuning}
\begin{figure}[ht]%
\centering
\includegraphics[width=1\textwidth]{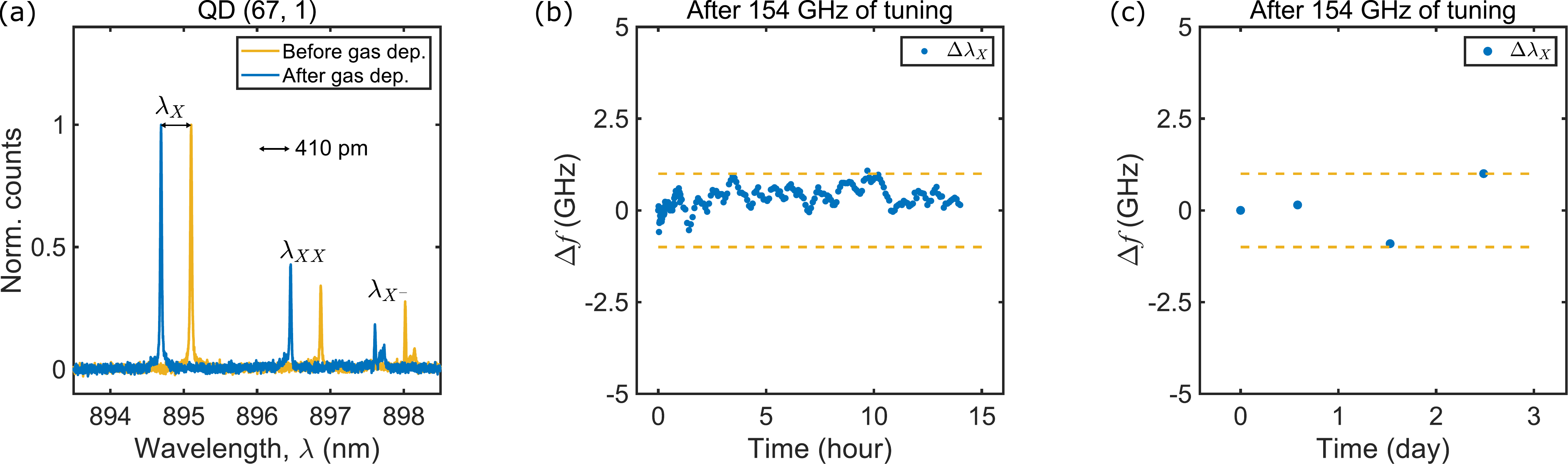}
\caption{Stability of the emission after the tuning: (a) The measurement is done by tuning the QD(67,1) emission by 410 pm or 154 GHz using the gas deposition method and constantly monitoring the emission spectrum afterwards. (b) The emission measured on the spectrometer for $\sim$15 hours. The dashed lines are representing the $\pm$1 GHz bounds. It is worth mentioning that the resolution of the spectrometer is on the order of GHz and we assume that most of the variations comes from the spectrometer resolution/stability rather than QD emission stability. (c) Same measurement like (b), however, for $\sim$3 days.
}
\label{fig: Supplemantary stability}
\end{figure}

\newpage
\section{Linewidth measurement \texorpdfstring{($X, XX, X^-$)}{X, XX, X-}}

\begin{figure}[h]%
\centering
\includegraphics[width=1\textwidth]{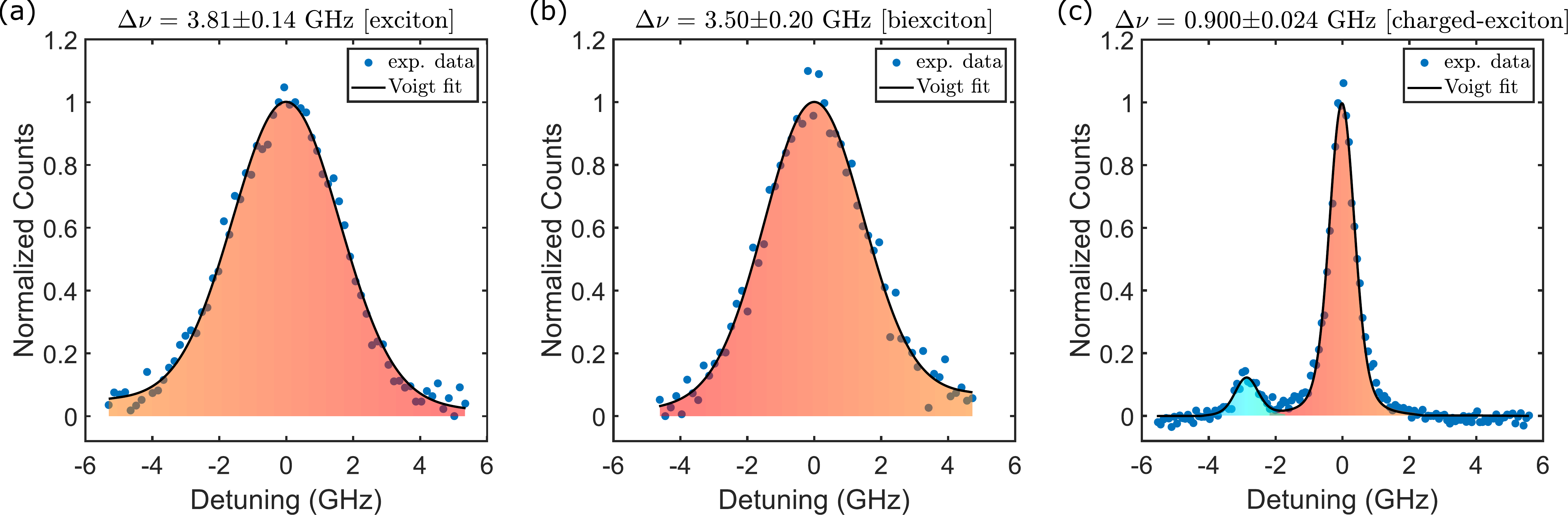}
\caption{Linewidth measurement: The measurement is done by sending the isolated QD emissions ($X$ or $XX$ or $X^-$) through a Fabry-P\'erot interferometer (10 GHz FSR, 67 MHz resolution) where one mirror of the cavity is fixed while the other mirror is scanned using a piezoelectric actuator by applying a variable voltage. The experimental data is fitted with a Voigt profile (see Method). (a) Exciton (b) Biexciton (c) negatively Charged exciton. As can be seen from the measurement, the charged exciton has $\sim$4-times smaller linewidth compared to the exciton and biexciton. Hence, the charged exciton emission is used for scanning the Cs D\textsubscript{1}-line for better resolution.
}
\label{fig: Supplemantary linewidth}
\end{figure}

\section{Analysis: Transmission through Cs vapour}

\begin{figure}[H]%
\centering
\includegraphics[width=1\textwidth]{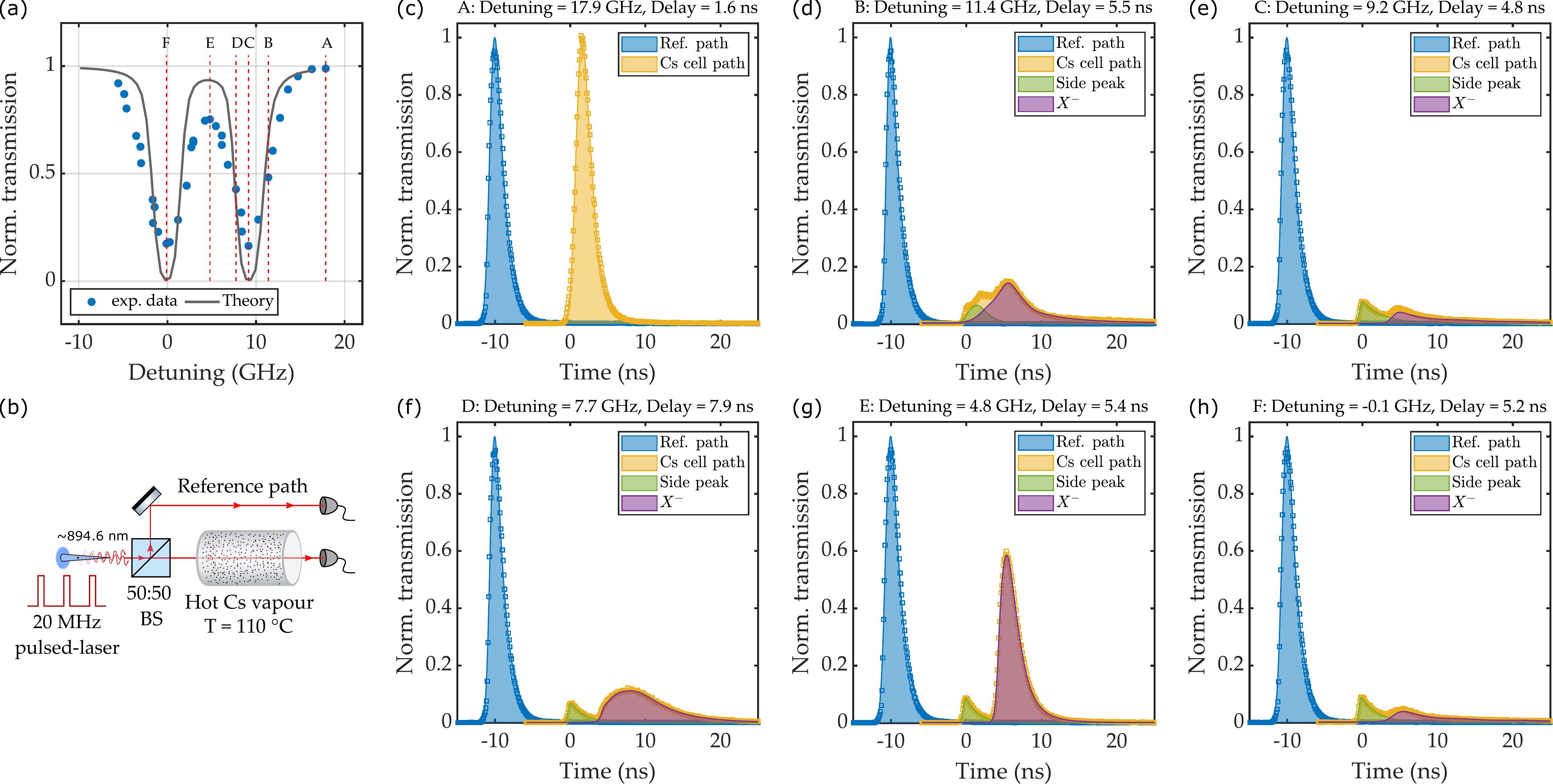}
\caption{ (a) Scanning of the QD $X^-$ emission through Cs D\textsubscript{1}-line. Here, normalized transmission is defined as the ratio of the total photons between the caesium cell path and the reference path. (b) Schematic of the experimental setup. An 80 ps long pulsed laser at 823 nm with a repetition rate of 20 MHz is used for above-bandgap excitation of the QD. $X^-$ photons from QD are sent to a 50:50 beam splitter, one path is going directly to the SNSPD and the other path is going through hot Cs vapour before the SNSPD. (c-h) Experimental raw data for 6 detuning locations, from A to F,  are depicted. After each pulsed-excitation of the QD, the arrival times of the emitted photons from the reference path and the Cs cell path are recorded and histogrammed. Blue and yellow regions are the histograms of the reference path and Cs cell path, respectively. The integration of an area is the total photon number into that region. When the QD emission is far detuned, point A in this case, it is assumed that the QD photons have negligible interactions with the Cs atoms. Hence, normalized transmission = area\textsubscript{Cs path}/area\textsubscript{ref. path} $\approx$ 1. There is a small delay of $\sim$1.6 ns measured compared to the room temperature transmission through the Cs cell. While the QD emission wavelength approaches near the resonances of the D\textsubscript{1}-line, the $X^-$ photons start getting absorbed and delayed. However, the $X_{side peak}$ emission is detuned by +63 pm (-24 GHz) from the $X^-$ wavelength (frequency) (Fig. \ref{fig: side_peak} in main text). Therefore, there is negligible effect from the Cs atoms on the $X_{side peak}$ emission compared to the $X^-$ emission. As can be seen on (d-h), the total photons through the cs path is illustrated by the yellow region, which is then decomposed into two regions. The green region is the $X_{side peak}$ photons and the purple region is the $X^-$ photons. The absorption and delay of the $X^-$ photons (purple) vary depending on the amount of detuning, while the $X_{side peak}$ photons (green) are almost unaltered. The decomposition is done by fitting the yellow region with 4 Lorentzians, one for the $X_{side peak}$ decay and the other three are to account for the oscillatory nature of the broadened $X^-$ pulse. The $X^-$ photons at the resonances are not completely absorbed, possibly due to the fact that: (i) broad linewidth ($\sim$900 MHz) of the $X^-$ photons, and (ii) $X_{side peak}$ photons are not completely eliminated in the fitting analysis.
}
\label{fig: Cs_cell_transmission_method}
\end{figure}

\newpage
\section{QD Photon -- Cs \texorpdfstring{D\textsubscript{1}}{D1} Line Propagation: Lorentzian Resonance and Density Matrix Model}

The field propagation is approximated via the slowly-varying envelope approximation. 
This is written as \citesupp{lukin2013modern}: 

\begin{linenomath} 
\begin{gather}
    \left( \frac{\partial}{\partial z} + \frac{1}{c} \frac{\partial}{\partial t} \right) \mathcal{E} (z, t) = \frac{i k}{2 \epsilon_0} \mathcal{P}(z, t) 
\end{gather}
\end{linenomath} 

The field and polarization envelopes are denoted by $\mathcal{E}$ and $\mathcal{P}$, respectively; $k$ is the free-space wave-vector of the field.  The field polarizes the atoms, which in turn, affects the field. 

\begin{figure}[ht]
\includegraphics[width=0.3\textwidth]{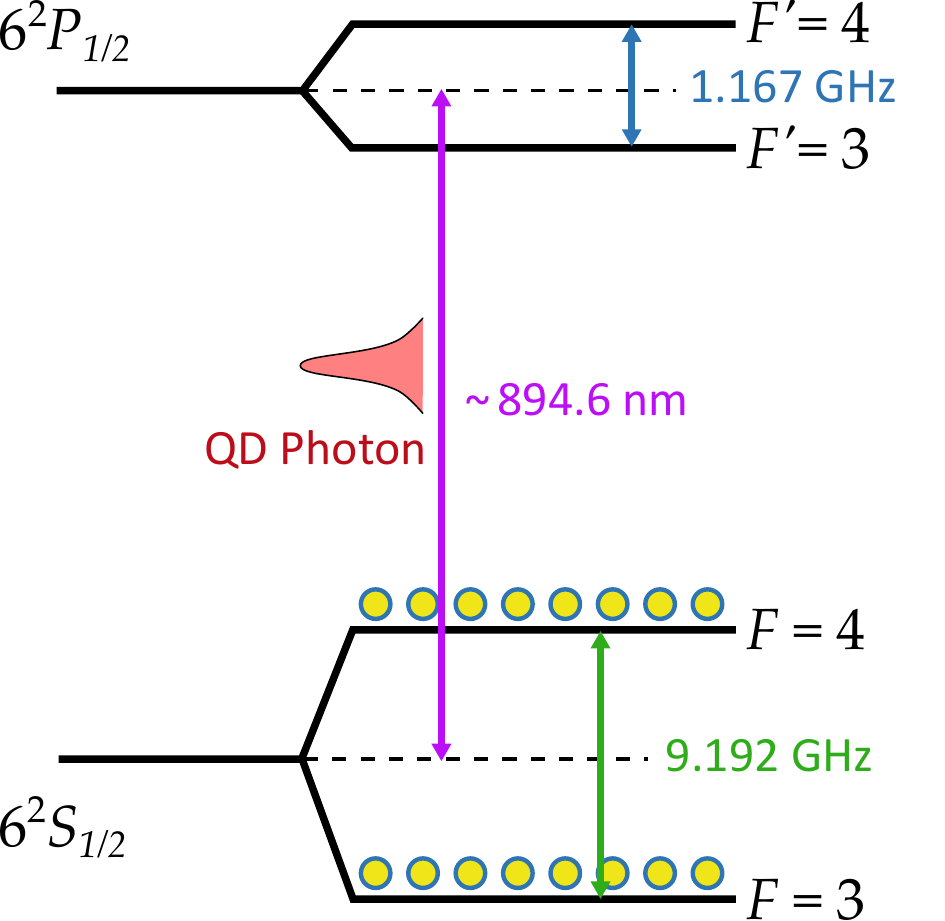}
\centering
\caption{A photon pulse interacting with the Cs D$_1$-line.}
\end{figure}

\section*{Lorentz Oscillator Model}

A two-level atom responds linearly (approximately) under continuous-wave weak-light excitation. Meaning that its polarization relates linearly to the field: $\mathcal{P} = \epsilon_0 \chi \mathcal{E}$. For a two-level atom (TLS), the susceptibility is given by a complex Lorentzian (assuming all the atoms are in the ground state) \citesupp{steck2003cesium}:
\begin{linenomath} 
\begin{gather*}
    \chi_{TLS} = i \frac{N}{V} \frac{\mu_{eg}^2}{\hbar \epsilon_0} \left( \frac{1}{\gamma - i \Delta} \right)
\end{gather*}
\end{linenomath}
The atomic transition is parameterized by: its transition dipole moment ($
\mu_{eg}$), decay rate ($\gamma$); detuning from the resonance ($\Delta$). The four transitions (Cs, D$_1$) from $F$ = 3, 4 $\longleftrightarrow$ $F'$ = 3, 4 form four Lorentzian resonances. The net susceptibility is given by \citesupp{camacho2007wide, akopian2011hybrid_Supp, vural2018two, kallmann1999propagation}:
\begin{linenomath} 
\begin{gather*}
    \chi_{Cs, \text{D}_1} = \sum \chi^{(j)} = i \frac{\rho}{\hbar \epsilon_0} \sum_j \frac{[\mu^{(j)}_{eg}]^2}{\gamma^{(j)} - i \Delta^{(j)}} 
\end{gather*}
\end{linenomath}
Fourier transforming Eq. (1), lends us the field after propagation:
\begin{linenomath} 
\begin{gather*}
    \frac{\partial \tilde{\mathcal{E}}(z, \nu)}{\partial z} =   \tilde{\mathcal{E}}(z, \nu) \left( \frac{ik \chi}{2} - \frac{i \nu}{c} \right)  \longrightarrow  \tilde{\mathcal{E}}(L, \nu) = \tilde{\mathcal{E}}(\nu, 0) \exp \left[i \left( \frac{k \chi(\nu)}{2} - \frac{\nu}{c} \right) L \right]
\end{gather*}
\end{linenomath}
Note that $\tilde{\mathcal{E}}(z, \nu) = \int \mathcal{E}(z, t) e^{- i \nu t} dt$ is frequency component ($\nu$) of the envelope. Inverse Fourier transforming the frequency-space field: $\tilde{\mathcal{E}}(z, \nu)$, provides the field in time: $\mathcal{E} (z, t)$.

\noindent The parameters describing the model are: \citesupp{steck2003cesium} (Detuning for a given $\nu$ is calculated from the C.O.M resonance and the hyperfine splittings. The transition dipole elements are estimated for polarized light incident on the vapour.)

\begin{table}[ht]
\centering
\begin{tabular}{|c|c|c|c|}
\hline
($F, F'$) &  $S_{F,F'}$ &  $\mu_{F,F'}$  &  $\Gamma_{F, F'}$  \\
\hline
(3,3) & 1/4 & $\sqrt{7/12}\mu_{\text{D}_1}$ & $S_{(3,3)}\Gamma_{\text{D}_1}$   \\
\hline
(3,4) & 3/4 & $\sqrt{7/4}\mu_{\text{D}_1}$ & $S_{(4,3)}\Gamma_{\text{D}_1}$   \\
\hline
(4,3) & 7/12 & $\sqrt{7/4}\mu_{\text{D}_1}$ & $S_{(3,4)}\Gamma_{\text{D}_1}$   \\
\hline
(4,4) & 5/12 & $\sqrt{5/4}\mu_{\text{D}_1}$ & $S_{(4,4)}\Gamma_{\text{D}_1}$  \\
\hline
\end{tabular}
\end{table}

\vspace{-9mm}

\begin{table}[ht]
\centering
\begin{tabular}{|c|c|}
\hline
$\mu_{\text{D}_1}$  &  $\Gamma_{\text{D}_1}$  \\
\hline
$3.1822 \times q_e a_0$ & $2\pi \times 4.5612 \text{MHz}$ \\
\hline
\end{tabular}
\end{table}
\vspace{-3mm}

\subsubsection*{Doppler Broadening}
For each frequency component of the pulse, $\nu$, the atomic response is broadened by the atomic velocity distribution. That is, $\Delta^{(j)} \rightarrow \Delta^{(j)} - k v_{m}$; where `$m$' is indexing the atoms -- which is Gaussian distributed ($\sigma_{DB}$) in one-dimension. 

\subsubsection*{QD Inhomogeneous Broadening}
The central frequency of the pulse ($\omega$) is Gaussian distributed ($\sigma^{f}_{IB}$) due to inhomogeneous broadening. The measured field intensity is averaged over intensities obtained for pulses with different central frequencies, weighted by the Guassian function. 

\noindent Below, $\sigma$ and $w$ are the standard deviation and beam waist of a Gaussian profile. These are found from the measured linewidth, given the width of a Lorentzian decay -- which is known from the lifetime $\tau$ as: $\tau$/(2$\pi$).
\begin{linenomath} 
\begin{gather*} 
    \text{FWHM}_{\text{Gaussian}} = 2 \sigma \sqrt{2 \text{ln}(2)} = 2 w \sqrt{\text{ln}(2)} \\
    \text{FWHM}_{\text{Voigt}} \approx \frac{\text{FWHM}_{\text{Lorentzian}}}{2} + \sqrt{\frac{\text{FWHM}^2_{\text{Lorentzian}}}{4} + \text{FWHM}^2_{\text{Gaussian}}}  
\end{gather*}
\end{linenomath} 

\subsubsection*{QD Above-Band Timing Jitter and Detector Response}
Lastly, the output time profile is convolved with two Gaussians in time: one, $\sigma^{t}_{TJ}$, to account for the timing jitter resulting from an above-band emission. The other, $\sigma^{t}_{DR}$, to account for the detector's response. 

\begin{table}[ht]
\centering
\begin{tabular}{|c|c|c|c|}
\hline
$\sigma_{DB}$ [MHz]  &  $\sigma^{f}_{IB}$ [MHz] & $\sigma^{t}_{TJ}$ [ns] & $\sigma^{t}_{DR}$ [ps] \\
\hline
300 - 500 \text{MHz} & 400 - 500 MHz & 1  & 200 \\
\hline
\end{tabular}
\end{table}

\section*{Density Matrix (Master Equation) Model}

More generally, the (averaged) polarization can be expressed in terms of the density matrix ($\rho$) as: $\langle \mathcal{P} \rangle = $Tr[$\hat{\mu} \rho$]. This can be followed via the Lindblad master equation. 
First, we consider one ground state and two excited states to see how this computation contrasts with the above model near the resonance (when the other 6\text{S}$_{1/2}$ state can be neglected). 

\noindent The ground state is labelled by `1', and the two excited states are labelled by `2' and `3'. The equations governing the field and atomic density matrix are \citesupp{cui2023}:
\begin{linenomath}
\begin{gather*}
    \left( \frac{\partial}{\partial z} + \frac{1}{c} \frac{\partial}{\partial t} \right) \mathcal{E}(z, t) = \frac{i k \rho}{2} [\mu_{13} \rho_{31} + \mu_{12} \rho_{21}] \longrightarrow \frac{\partial}{\partial z} \mathcal{E}(z, \tau) = \frac{i k \rho}{2} [\mu_{13} \rho_{31} + \mu_{12} \rho_{21}] \\
    \frac{\partial \rho}{\partial \tau} = -\frac{i}{\hbar}[\hat{H}, \rho] + \sum_{i=2,3} \gamma_i (\hat{\sigma}_{1i} \rho \hat{\sigma}_{i1} - \hat{\sigma}_{ii} \rho - \rho \hat{\sigma}_{ii})
\end{gather*}
\end{linenomath} 

\noindent $\hat{\sigma}_{ij}$ represents the atomic operator associated with the levels $i$, $j$. A semi-implicit finite difference method is implemented to evaluate the coupled equations, yielding pulse propagation (and density matrix evolution terms) \citesupp{drummond1983central}.

\renewcommand{\bibnumfmt}[1]{[S#1]}
\bibliographystylesupp{sn-standardnature}
\bibliographysupp{2bibliography}

\end{document}